\documentclass[12pt]{article}
\usepackage{latexsym,graphicx,a4,epsfig,psfrag,here}    

\newcommand{\bda}{\begin{\displaymath}\begin{array}{rl}}
\newcommand{\eda}{\end{array}\end{displaymath}}
\newcommand{\be}{\begin{equation}}
\newcommand{\ee}{\end{equation}}
\newcommand{\bdm}{\begin{displaymath}}
\newcommand{\edm}{\end{displaymath}}
\newcommand{\bea}{\begin{eqnarray}}
\newcommand{\eea}{\end{eqnarray}}
\newcommand{\no}{\nonumber \\}
\newcommand{\fs}{\; \; .}
\newcommand{\co}{\; \; ,}
\newcommand{\al}{&\!\!\!\!}
\newcommand{\eff}{{e\hspace{-0.1em}f\hspace{-0.18em}f}}
\newcommand{\ind}{\scriptscriptstyle}

\newcommand{\ubar}{\overline{\rule[0.42em]{0.4em}{0em}}\hspace{-0.5em}u}

\newcommand{\lbar}{\bar{\ell}}

\newcommand{\Wbar}{\,\overline{\rule[0.75em]{0.9em}{0em}}\hspace{-1em}W}
\newcommand{\Pbar}{\,\overline{\rule[0.75em]{0.5em}{0em}}\hspace{-0.7em}P}

\newcommand{\pbar}{\overline{\rule[0.5em]{0.4em}{0em}}\hspace{-0.5em}p}

\newcommand{\lvac}{\langle 0|\,}
\newcommand{\rvac}{\,|0\rangle}

\newcommand{\rs}{\langle r^2\rangle\rule[-0.2em]{0em}{0em}_s}
\begin{document}
\begin{titlepage}

\begin{flushright}ZU-TH 11/00\\
BUTP-00/17\end{flushright}

\vspace{3cm}
\begin{center}{\Large\bf The $\pi\pi$ $S$-wave scattering lengths}

\vspace{0.5cm}
July 24, 2000

\vspace{0.5cm}
G.~Colangelo$^a$,
J.~Gasser$^b$ and  
H.~Leutwyler$^b$
\vspace{2em}
\footnotesize{\begin{tabular}{c}
$^a\,$Institute for Theoretical Physics, University of 
Z\"urich\\
Winterthurerstr. 190, CH-8057 Z\"urich, Switzerland\\
$^b\,$Institute for Theoretical Physics, University of 
Bern\\   
Sidlerstr. 5, CH-3012 Bern, Switzerland
\end{tabular}  }

\vspace{1cm}

\begin{abstract}

We match the known chiral perturbation theory representation of 
the $\pi\pi$ scattering amplitude to two loops
with a phenomenological description that relies on the Roy equations. 
On this basis, the corrections to Weinberg's low energy theorems for the 
$S$-wave scattering lengths are worked out to second order in the
expansion in powers of the quark masses. The resulting predictions, 
$a_0^0=0.220\pm 0.005$, $a_0^2=-0.0444\pm 0.0010$, contain remarkably small 
uncertainties and thus allow a very sensitive
experimental test of the hypothesis that the quark condensate is the leading 
order parameter of the spontaneously broken chiral symmetry.  
\end{abstract}

\vspace{4cm}
\footnotesize{\begin{tabular}{ll}
{\bf{Pacs:}}$\!\!\!\!$& 11.30.Rd, 11.55.Fv, 11.80.Et, 12.39.Fe, 13.75.Lb\\
{\bf{Keywords:}}$\!\!\!\!$& Roy equations, Dispersion relations, 
Partial wave analysis,\\&
Meson-meson interactions, Pion-pion scattering, Chiral symmetries,\\
&Chiral Perturbation Theory
\end{tabular}} 
\end{center}
\end{titlepage}

\setcounter{page}{2}

\subsubsection*{1. Introduction}
In the chiral limit, where the masses of the two lightest quarks are turned
off, QCD acquires an exact SU(2)$\times$SU(2) symmetry. We rely on the 
standard picture, where it is assumed that this symmetry is broken 
spontaneously and that the quark condensate $\lvac \ubar u\rvac$ represents 
the leading order parameter. The quark masses act as symmetry
breaking parameters, which equip the Goldstone bosons -- the pions -- with a 
mass\footnote{We disregard isospin breaking and
set $m_u=m_d=m$. In the numerical work, we identify $M_\pi$ with the mass
of the charged pion and use $F_\pi=92.4\,\mbox{MeV}$.}: 
$M_\pi^2=2 m B+O(m^2)$, where $B$ stands for the value of
$|\lvac \ubar u \rvac|/F_\pi^2$ in the chiral limit. 

Since Goldstone bosons
can interact only if they carry momentum, the $\pi\pi$ $S$-wave
scattering lengths vanish in the chiral limit. Hence these quantities
represent a sensitive probe of the symmetry breaking generated by
the quark masses. Weinberg's low energy theorems \cite{Weinberg 1966}
state that their values are related to the pion mass, which
also represents a symmetry breaking effect: 
\bea\label{WLET} a_0^0=\frac{7\,M_\pi^2}{32\,\pi\,F_\pi^2}+O(m^2)\co
\hspace{2em} a_0^2=-\frac{M_\pi^2}{16\,\pi\,F_\pi^2}+O(m^2)\fs
\eea
The relations hold to leading order in the expansion in powers of $m$.
The corrections of order $m^2$ were worked out in \cite{GL 1983}. As an
example, we quote the result for the combination $2a_0^0-5a_0^2$, which is
particularly simple:
\bea\label{a02 one loop}
2a_0^0-5a_0^2=\frac{3M_\pi^2}{4\pi F_\pi^2}\left\{1+\mbox{$\frac{1}{3}$}
M_\pi^2\rs+\mbox{$\frac{41}{12}$}\xi \right\}+O(m^3)\fs\eea
$\rs$ is the mean square radius of the scalar form factor
and $\xi$ measures the pion mass in units of the
scale set by $4\pi F_\pi\simeq  1\,\mbox{GeV}$:
\bea \xi=\left(\frac{M_\pi}{4\pi F_\pi}\right)^{\!2}=0.01445\fs\eea
With the evaluation of the chiral perturbation series to two loops
described in ref.~\cite{BCEGS}, the low energy expansion of the scattering
amplitude is now known to next-to-next-to-leading order. The purpose of the 
present paper is to analyze the consequences for the scattering lengths.

\subsubsection*{2. Chiral representation of the scattering amplitude}
The two-loop representation yields the first three terms in the low energy
expansion of the partial waves:
\be\label{seriespw} t^I_\ell(s)=t^I_\ell(s)_2+ t^I_\ell(s)_4+
t^I_\ell(s)_6+O(p^8)\fs \ee
Since inelastic reactions start showing up only at
$O(p^8)$, unitarity implies
\bea \mbox{Im}\,t^I_\ell(s)=\sigma(s)\,|\hspace{0.04em}t^{I}_\ell(s)
\hspace{0.03em}|^2+O(p^8)\co\hspace{2em}
\sigma(s)= (1-4M_\pi^2/s)^{\frac{1}{2}}
\fs\eea
The condition fixes the imaginary parts of the two-loop
amplitude in terms of the one-loop representation. At leading order, 
the scattering amplitude is linear in the Mandelstam variables, so that
only the S- and P-waves are different from zero. Unitarity therefore
implies that, up to and including $O(p^6)$, only these partial waves
develop an imaginary part. Accordingly, the chiral representation 
of the scattering amplitude can be written in the form \cite{Knecht Moussallam
  Stern Fuchs}   
\bea\label{chiral decomposition} A(s,t,u)
\al=\al C(s,t,u)+32\pi\left\{\mbox{$\frac{1}{3}$}\,U^0(s)+
\mbox{$\frac{3}{2}$}\,(s-u)\,U^1(t)
+\mbox{$\frac{3}{2}$}\,(s-t)\,U^1(u)\right.\no
\al\al \left.+\mbox{$\frac{1}{2}$}\,U^2(t)+\mbox{$\frac{1}{2}$}\,U^2(u)
-\mbox{$\frac{1}{3}$}\, U^2(s) \right\}+ O(p^8)\co\eea
where $C(s,t,u)$ is a crossing symmetric polynomial,
\be\label{poly}
C(s,t,u)=c_1+s\,c_2+s^2\,c_3
+(t-u)^2\,c_4+s^3\,c_5
+s\,(t-u)^2\,c_6\fs \ee
The functions $U^0(s)$, $U^1(s)$ and $U^2(s)$ describe the 
``unitarity corrections'' associated with $s$-channel isospin $I=0,1,2$,
respectively. In view of $\mbox{Im}\,t^I_\ell(s)_6\propto s^3$, several
subtractions are needed for the dispersive representation of these functions
to converge. We subtract at $s=0$ and set
\bea\label{disp U}U^I(s)\al=\al \frac{s^{4-\epsilon_I}}{\pi}
\int_{4M_\pi^2}^\infty ds'\;\frac{\sigma(s')\,t^I(s')_2\,\{t^I(s')_2+ 2\,
\mbox{Re} \,t^I(s')_4\} }
{s^{\prime\,4-\epsilon_I}(s'-4M_\pi^2)^{\epsilon_I}(s'-s )}\fs\eea
The subtraction constants are collected in $C(s,t,u)$.
As only the $S$- and $P$-waves enter, we have dropped the lower index,
$\{t^0,t^1,t^2\}=\{t^0_0,t^1_1,t^2_0\}$. For kinematic reasons, the
integrand of the $P$-wave differs from the one of the $S$-waves:
$\{\epsilon_0,\epsilon_1,\epsilon_2\}=\{0,1,0\}$.
It is straightforward to check that the result of the two-loop 
calculation \cite{BCEGS} is indeed of this structure. 

\subsubsection*{3. Effective coupling constants}
In the following, the two-loop result for the polynomial part of the
amplitude plays a key role. It involves the coupling constants occurring in
the derivative expansion of the effective Lagrangian, ${\cal
L}_{\eff}={\cal L}_2+{\cal L}_4+{\cal L}_6+\ldots$ The corresponding
formulae, which specify how the coefficients $c_1,\ldots,\,c_6$ depend on
the quark masses, can be worked out from the two-loop representation given
in ref.~\cite{BCEGS} (for explicit expressions, see \cite{CGL}). These
formulae, in particular contain Weinberg's low energy theorems, which in
this language state that the leading terms in the expansion of the first
two coefficients are fixed by $M_\pi$ and $F_\pi$:
$c_1=-M_\pi^2/F_\pi^2+\ldots\,$, $c_2=1/ F_\pi^2+\ldots$ At first order,
the constants $\ell_1,\ell_2,\ell_3,\ell_4$ from ${\cal L}_4$ \cite{GL
1984} enter, and at second order, the chiral representation of the
scattering amplitude involves the couplings $r_1,\ldots\,,r_6$ from ${\cal
L}_6$ \cite{BCE}.  We need to distinguish two categories of coupling
constants:
\begin{description}\item {\it a. Terms that survive in
the chiral limit.} Four of the coupling constants that enter the two-loop
representation of the scattering amplitude belong to this category:
$\ell_1,\,\ell_2,\,r_5,\,r_6$.
\item {\it b. Symmetry breaking terms.}
The corresponding vertices are proportional to a power
of the quark mass and involve the coupling constants $\ell_3$, $\ell_4$,
$r_1$, $r_2$, $r_3$, $r_4$.\end{description}
The constants of the first category show up in the momentum dependence of 
the scattering amplitude, so that these couplings may be determined
phenomenologically. The symmetry breaking terms, on the other hand,
specify the dependence of the amplitude on the quark masses. Since these 
cannot be varied experimentally, information concerning the second category
of coupling constants can only be obtained from sources other than $\pi\pi$
scattering. The constants $r_n$ from ${\cal L}_6$
only generate tiny effects, so that crude theoretical estimates
suffice, but the couplings $\ell_3$ and $\ell_4$ from ${\cal L}_4$
do play an important role and we now briefly discuss the information that
we will be using for these. 

The crucial parameter that distinguishes the standard framework from the 
one proposed in ref.~\cite{Stern Sazdjian Fuchs} 
is $\ell_3$. This coupling constant determines the 
first order correction in the Gell-Mann-Oakes-Renner-relation:
$ M_\pi^2= 2 B m\left\{1-\frac{1}{2}\xi\lbar_3
+O(\xi^2)\right\}$.
The value of $\ell_3$ is not known accurately.
Numerically, however, a significant change in the prediction for the 
scattering lengths can only arise if the crude estimate in 
ref.~\cite{GL 1984},
\be\label{l3barnum} \bar{\ell}_3=2.9\pm 2.4\,,\ee 
should turn out to be entirely wrong. We do not make an attempt
at reducing the uncertainty in $\ell_3$ within the
standard framework, but will 
explicitly indicate the sensitivity to this coupling constant. 

Chiral symmetry implies that the coupling constant $\ell_4$ also shows up in 
the expansion of the scalar radius in powers of the quark
masses \cite{GL 1983}: 
\bea \rs = \frac{3}{8\pi^2 F_\pi^2}
\left\{\lbar_4-\frac{13}{12}+\xi\,\Delta_r+O(\xi^2)\right\}\fs
\eea
As pointed out in ref.~\cite{DGL}, the scalar radius can be determined
on the basis of a dispersive evaluation of the scalar form factor.
We have repeated that calculation with the information about the
phase shift $\delta_0^0(s)$ obtained in ref.~\cite{ACGL}.
In view of the strong final state interaction in the 
$S$-wave, the scalar radius is significantly larger than the electromagnetic 
one, $\langle r^2\rangle_{\ind  e.m.}= 0.439\pm 0.008 \,\mbox{fm}^2$ 
\cite{Amendolia}. The result,
\be\label{rsnum} \rs=0.61\pm 0.04\,\mbox{fm}^2\co\ee 
confirms the value given in ref.~\cite{DGL} and is consistent with earlier 
estimates of the low energy constant $\ell_4$ based on the
symmetry breaking seen in $F_K/F_\pi$ or on the decay $K\rightarrow
\pi\ell\nu$ \cite{GL 1985}, but is considerably 
more accurate. Since the chiral representation of the scalar form factor is
known to two loops 
\cite{Bijnens Colangelo Talavera}, the dependence of the correction
$\Delta_r$ on the quark masses is also known. In
addition to $\ell_1,\ldots\,,\ell_4$, the explicit expression involves
a further term, $r_{S2}$, from ${\cal L}_6$ \cite{BCE}. In the following,
we use this representation to eliminate the parameter $\ell_4$ in
favour of the scalar radius.

\subsubsection*{4. Low energy theorems}
We now show that chiral symmetry implies two relations among the coefficients
$c_1,\ldots\,,c_4$. For this purpose we consider the combinations
\bea \label{defC12}
\hspace*{-6em}C_1\al\equiv \al F_\pi^2\left\{c_2+4M_\pi^2(c_3-c_4)\right\}
\,,\hspace{0.5em}C_2\equiv \frac{F_\pi^2}{M_\pi^2}
\left\{-c_1+4M_\pi^4(c_3-c_4)\right\}\,.\eea
Chiral symmetry implies that, if the quark masses are turned off, both
$C_1$ and $C_2$ tend to 1. The contributions from $c_3$ and $c_4$ ensure that
the first order corrections only involve the symmetry breaking couplings
$\ell_3$ and $\ell_4$. Eliminating $\ell_4$ in favour of the scalar radius,
the low energy theorems take the form
\bea\label{letcr} 
C_1 \al=\al 1+
\frac{M_\pi^2}{3}\,\rs+\frac{23\,\xi}{420}+\xi^2 \Delta_1+O(\xi^3)\co\\
C_2\al=\al
1+\frac{M_\pi^2}{3}\,\rs+ \frac{\xi}{2}\left\{\lbar_3-
\frac{17}{21}\right\} +\xi^2 \Delta_2+O(\xi^3)\nonumber \fs\eea
At first nonleading order, $C_1$ is fully determined by 
the contribution from the scalar radius, while $C_2$ also contains a 
contribution from $\ell_3$.
Inserting the values $\rs=0.61 \,\mbox{fm}^2$ and $\lbar_3=2.9$ and ignoring 
the two-loop corrections $\Delta_1$, $\Delta_2$, we obtain
$C_1=1.103$, $C_2=1.117$. The value of $C_2$
differs little from $C_1$ -- as stated above, 
the estimate (\ref{l3barnum}) implies that the contributions from 
$\ell_3$ are very small. The size of the two-loop corrections will be 
discussed later on, when the phenomenological information about
the coefficients $c_1,\ldots\,,c_4$ has been sorted out.

\subsubsection*{5. Phenomenological representation of the scattering amplitude}
Our predictions for the scattering lengths are based on the comparison of the
chiral representation with the one that follows from analyticity and
unitarity alone. As shown by Roy \cite{Roy}, the fixed-$t$ dispersion
relations can be written in such a form that they express the $\pi\pi$
scattering amplitude
in terms of the imaginary parts in the physical region of the $s$-channel.
The resulting representation for $A(s,t,u)$ contains 
two subtraction constants, which may be identified with the scattering
lengths $a^0_0$ and $a^2_0$. Unitarity converts this representation
into a set of coupled integral equations, which we recently examined
in detail \cite{ACGL}. The upshot of that analysis is that
$a_0^0$ and $a_0^2$ are the essential low energy parameters: Once these are
known, the available experimental data determine the
behaviour of the $\pi\pi$ scattering amplitude at low energies to within
remarkably small uncertainties. In the present context, the main
result of interest is that the representation allows us to determine the 
imaginary parts of the partial waves in terms of $a_0^0$ and
$a_0^2$. Since the resulting representation is based on the available 
experimental information, we refer to it as the phenomenological
representation. 

The branch cut generated by the imaginary parts of the partial waves with 
$\ell\geq2$ starts manifesting itself only at $O(p^8)$. Accordingly, we may
expand the corresponding contributions to the dispersion integrals
into a Taylor series of the momenta. The singularities due to the imaginary
parts of the $S$- and $P$-waves, on the other hand, show up
already at $O(p^4)$ -- these cannot be replaced
by a polynomial. The corresponding contributions to the amplitude
are of the same structure as the unitarity corrections and also involve
three functions of a single variable. We subtract the relevant dispersion 
integrals in the same manner as for the chiral representation:
\bea\label{dispWbar}\Wbar^I(s)\al=\al 
\frac{s^{4-\epsilon_I}}{\pi}\int_{4M_\pi^2}^{\infty}
ds'\;\frac{\mbox{Im}\,t^I(s')}
{s^{\prime\,4-\epsilon_I}(s-4M_\pi^2)^{\epsilon_I}(s'-s )}\fs\eea
Since all other contributions can be replaced by a polynomial, the
phenomenological amplitude takes the form
\bea\label{phenrep} A(s,t,u)\al=\al 16\pi a_0^2+
\frac{4\pi }{3M_\pi^2}\,(2a_0^0-5a_0^2)\,s+
\Pbar(s,t,u) \\\al\al +32\pi\left\{\mbox{$\frac{1}{3}$}
\Wbar^0(s)+\mbox{$\frac{3}{2}$}(s-u)\Wbar^1(t)
+\mbox{$\frac{3}{2}$}(s-t)\Wbar^1(u)\right.\no
\al\al \left.\hspace{2.3em}+\mbox{$\frac{1}{2}$}\Wbar^2(t)+
\mbox{$\frac{1}{2}$}\Wbar^2(u)
-\mbox{$\frac{1}{3}$} \Wbar^2(s) \right\}+O(p^8)\,.\nonumber\eea
We have explicitly displayed the contributions from the subtraction 
constants $a_0^0$ and $a_0^2$.
The term $\Pbar(s,t,u)$ is a crossing symmetric polynomial
\bea \Pbar(s,t,u)\al=\al \pbar_1+\pbar_2\,s+\pbar_3\,s^2+\pbar_4\,(t-u)^2+
\pbar_5\,s^3+\pbar_6\,s(t-u)^2\fs\eea
Its coefficients can be expressed in terms of integrals over the imaginary
parts of the partial waves. We do not list the explicit expressions here, but 
refer to \cite{CGL}. In the following, the essential point is that the
coefficients $\pbar_1,\ldots\,,\,\pbar_6$ can be determined phenomenologically.

\subsubsection*{6. Matching the two representations}
We now show that, in their common domain of validity, the two 
representations of the scattering amplitude specified above agree,
provided the parameters occurring therein are properly matched.
The key observation is that, in the integrals (\ref{dispWbar}), 
only the region where $s'$ is of order $p^2$ matters for
the comparison of the two representations. The remainder generates 
contributions to the amplitude that are most of order $p^8$. 
Moreover, for small values of $s'$, the quantities $\mbox{Im}\,t^I_\ell(s')$
are given by the one-loop representation, except for contributions 
that again only manifest themselves at $O(p^8)$. This implies that
the differences between the functions $\Wbar^I(s)$ and $U^I(s)$
are beyond the accuracy of the two-loop representation  \cite{Knecht
  Moussallam Stern Fuchs}.  
Hence the two descriptions agree if and only if the polynomial parts
do,
\bea \label{mc}C(s,t,u)=16\pi a_0^2+
\frac{4\pi }{3M_\pi^2}\,(2a_0^0-5a_0^2)\,s+\Pbar(s,t,u)+O(p^8)\fs
\eea 
Since the main uncertainties in the coefficients of the 
polynomial $\Pbar(s,t,u)$ arise from their sensitivity to
the scattering lengths $a^0_0$, $a^2_0$,
the above relations essentially determine the coefficients $c_1,\ldots,\,c_6$
in terms of these two observables. The same then also holds for the 
quantities $C_1$, $C_2$ defined in eq.~(\ref{defC12}).
The corresponding low energy theorems for $a_0^0$ and $a_0^2$ are of the form
\bea \label{aC}
a^0_0\al =\al \frac{7M_\pi^2C_0}{32\pi F_\pi^2} +M_\pi^4\,\alpha_0
+O(m^4)\,,\hspace{0.5em}
a^2_0=- \frac{M_\pi^2 C_2}{16\pi F_\pi^2} +M_\pi^4\,\alpha_2
+O(m^4)\,,
\eea
with $C_0\equiv\frac{1}{7}(12\, C_1-5 \,C_2)$. The terms 
$\alpha_0$, $\alpha_2$ stand for integrals over the imaginary parts of
the partial waves that can be worked out from the available experimental
information. Since the behaviour of the
imaginary parts near threshold is sensitive to the scattering lengths we
are looking for, the same applies to these integrals. In the narrow range of
interest, the dependence is very well described by 
\bea  M_\pi^4\alpha_0\al=\al .04478+.30 \Delta a_0^0-.37 
\Delta a_0^2
+ .5(\Delta a_0^0)^2-1.2\Delta a_0^0\Delta a_0^2+1.8(\Delta a_0^2)^2\,,\no 
 M_\pi^4\alpha_2\al=\al
.0055+.023 \Delta a_0^0-.095\Delta a_0^2 + .01(\Delta a_0^0)^2
-.12\Delta a_0^0 \Delta a_0^2+.66(\Delta a_0^2)^2 
\,,\nonumber\eea
with $\Delta a_0^0\equiv a_0^0-0.225$, $\Delta a_0^2\equiv a_0^2+0.03706$.
\subsubsection*{7. Results for $a_0^0$ and $a_0^2$}

The representation (\ref{aC}) splits the correction to Weinberg's leading 
order formulae
into two parts: a correction factor $C_n$, which at first nonleading order
only involves the scalar radius and the coupling constant $\ell_3$,
and a term $\alpha_n$ that can be determined on phenomenological grounds. 

Inserting the one-loop prediction for $C_1$, $C_2$ 
in the relations (\ref{aC}) and solving for
$a_0^0,a_0^2$, we obtain the following first order results:
\bea\label{one loop} a_0^0    =  0.2195\,,\hspace{0.5em}a^2_0 = -0.0446\,,
\hspace{0.5em}
2a_0^2-5a^2_0=0.662\,.\eea

The two-loop corrections $\Delta_1$ and $\Delta_2$ involve the
coupling constants $\ell_1,\,\ell_2,\,\ell_3$, the scalar radius, as well
as the terms $r_1,\ldots\,,r_4$, $r_{S2}$ from ${\cal L}_6$. 
The size of the contributions from the latter
may be estimated with the resonance model described in
refs.~\cite{BCEGS,Bijnens Colangelo Talavera}.
The constants $\ell_1$, $\ell_2$ can then be determined numerically with the 
phenomenological values of $c_3$ and $c_4$. 
The resulting two-loop corrections for the scattering lengths are very small.
The numerical result is sensitive to the value of the 
scale $\mu$ at which the renormalized coupling constants $r^r_n(\mu)$ are
assumed to be saturated by the resonance contributions. For
$500\,\mbox{MeV}<\mu <1\,\mbox{GeV}$, the corrections vary in the range 
$-0.001 < \xi^2\Delta_1< 0.003$, $\,-0.004 < \xi^2\Delta_2< 0.001$. 
In the following, we use the resonance model at the scale 
$\mu=M_\rho$ and take the above range as an estimate for the uncertainties to
be attached to the two-loop corrections.

To estimate the errors due to the phenomenological input,
we vary the imaginary parts within the range discussed in ref.~\cite{ACGL},
where the Roy equations are used to determine the behaviour of the $S$- and
$P$-waves below 800 MeV. If the $S$-wave scattering lengths are held fixed,
the variations in the values of $\alpha_0$, $\alpha_2$ are dominated by 
those in the input used for the 
phases at 800 MeV. The corresponding uncertainty
in $\{a_0^0,\,a_0^2,\,2a_0^0-5a_0^2\}$ amounts to 
$\{\pm .9,\,\pm .2,\,\pm 1\}\cdot 10^{-3}$. 

We conclude that the uncertainties are dominated by those in $\ell_3$ and 
$\rs$. Adding the remaining sources of error up, we obtain\footnote{
The tiny errors given in eq.(\ref{l3rdependence}) merely 
indicate the noise seen in our calculation --
we do not claim to describe the scattering amplitude to that accuracy
(compare section 14.1 of \cite{ACGL}).}
\bea\label{l3rdependence} a_0^0\al=\al 0.220\pm 0.001
 -0.0017\,\Delta \ell_3+ 0.027\, \Delta_{r^2}\,,\\
a^2_0\al=\al -0.0444 \pm 0.0003 
-0.0004\,\Delta \ell_3-0.004\, \Delta_{r^2}\,,
\nonumber\eea
with $\lbar_3=2.9+\Delta \ell_3$, $\rs=0.61\,\mbox{fm}^2(1+\Delta_{r^2})$. 
Inserting the estimates (\ref{l3barnum}), (\ref{rsnum}), we arrive at our 
final result:
\bea\label{final result} \begin{array}{ll}
a_0^0= 0.220\pm 0.005\,,\al a^2_0=-0.0444\pm
0.0010\,,\\
2a_0^0-5a_0^2= 0.663\pm 0.006\,,\hspace{2em}\al a^0_0-a^2_0= 0.265\pm0.004\,.
\end{array}
\eea

\subsubsection*{8. Discussion}
The truncation of the chiral perturbation series represents an inherent
limitation of our calculation. We have shown, however, that the 
corrections of $O(p^6)$ barely change the predictions for $a_0^0$ and $a_0^2$
obtained at $O(p^4)$. For this reason, we expect the contributions from yet
higher orders to be entirely negligible. 

The rapid convergence of the series is a virtue of the specific method used
to match the chiral and phenomenological representations. To demonstrate
this, we briefly discuss the alternative approach used in ref.~\cite{GL
1983,BCEGS}, where the results for the various scattering lengths and
effective ranges are obtained by directly evaluating the chiral
representation of the scattering amplitude at threshold. We instead express
the amplitude in terms of three functions of a single variable $s$ and
match the coefficients of the Taylor expansion at $s=0$ -- in this
language, the approach of ref.~\cite{GL 1983,BCEGS} amounts to a matching
at threshold. It is straightforward to work out the chiral representation
at threshold with the values of the effective coupling constants that we
find with our method. Truncating the series at $O(p^4)$, we obtain
$\{a_0^0,a_0^2,2a_0^0-5a_0^2\}=\{0.200,-0.0445,0.624\}$, in agreement with
the values\footnote{As stated in \cite{GL 1983}, the error bars given only
measure the accuracy to which the first order corrections can be
calculated; they do not include an estimate of contributions due to higher
order terms. The small numerical differences arise partly from the manner
in which the coupling constants $\lbar_1$, $\lbar_2$ are determined, partly
from the values used for $F_\pi$ -- the old number, 93.3 MeV, does not
account for radiative corrections.} of ref.~\cite{GL 1983}: $\{0.20\pm
0.01,-0.042\pm0.002,0.614 \pm0.018\}$. At threshold, the terms of $O(p^6)$
are by no means negligible: They take the values obtained at $O(p^4)$ into
$\{0.215,-0.0445,0.653\}$, in agreement with the results of
ref.~\cite{BCEGS,Kl4 two loops, Nieves:1999zb}. These numbers describe the
expansion of the scattering lengths in powers of the quark
masses\footnote{More precisely, the expansion parameter is the physical
pion mass and, in lieu of the coupling constant $F$, the physical decay
constant is held fixed.} to order $m^3$. In the case of $a_0^0$, for
instance, the numerical values to order $m$, $m^2$ and $m^3$ are
$a_0^0=0.159$, $0.200$ and $0.215$, respectively -- these correspond to the
diamonds shown in the figure.

The reason why the straightforward expansion of the scattering lengths in
powers of the quark masses converges rather slowly is that these represent
the values of the amplitude at threshold, that is at the place where the 
branch cut required by unitarity starts. 
The truncated chiral representation does not describe that 
singularity well enough, particularly at one loop, where the relevant
imaginary parts stem from the tree level approximation. 
The matching must be done in such a manner that
the higher order effects are small. In contrast to a matching at threshold --
that is, to the straightforward
expansion of the scattering lengths --
our method fulfills this criterion remarkably well: We are using the expansion
in powers of the quark masses only for the coefficients $C_1$ and $C_2$, while
the curvature generated by the unitarity cut is evaluated phenomenologically.
Solving eq.(\ref{aC}) with the expansion of these coefficients to order
$1$, $m$ and $m^2$, we obtain a much more rapidly convergent series: 
$a_0^0=0.197$, $0.2195$, $0.220$.

If the effective coupling constants are the same, the only 
difference between the two approaches is the one between the 
functions $\Wbar^I(s)$ and $U^I(s)$. In particular, the results 
for $a_0^0$, $a_0^2$ only differ  
because the numerical values of 
$\Wbar^I(s)$ and $U^I(s)$ at $s=4M_\pi^2$ are not the same.
As mentioned above, the difference between the two sets of functions affects 
the scattering amplitude only at $O(p^8)$ and beyond. Numerically,
however, it is not irrelevant which one of the two is used to describe the
effects generated by the unitarity cuts: While the functions $\Wbar^I(s)$
account for the imaginary parts of the $S$- and $P$-waves to the accuracy to
which these are known, the quantities $U^I(s)$ represent a comparatively
crude approximation, obtained by evaluating the imaginary parts
with the one-loop representation. 

\subsubsection*{9. Conclusion}

Our analysis relies on two ingredients: the evaluation of the chiral
perturbation series for the $\pi\pi$ scattering amplitude 
to two loops \cite{BCEGS} and the phenomenological representation obtained
in ref.~\cite{ACGL} by solving the Roy equations.
We have shown that the comparison of the two descriptions
allows us to predict the $S$-wave scattering lengths at the 2--3\% level
of accuracy.

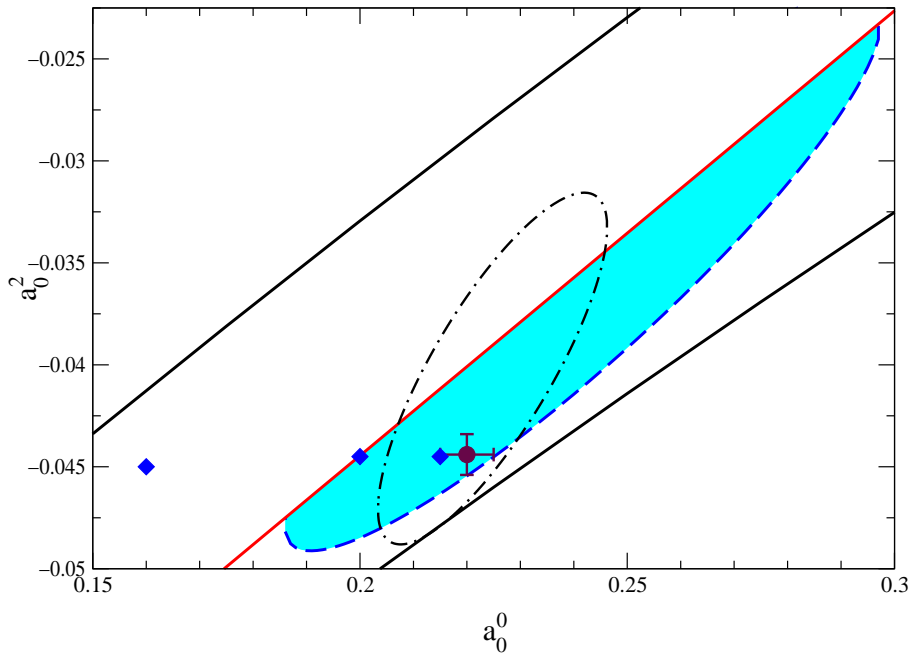
\vspace*{-1em}
\begin{figure}[thb] 
\leavevmode \begin{center}
\includegraphics[angle=-90,width=12cm]{allowed_region2}
\end{center}
\end{figure}

\vspace*{-2em}
\noindent
{\small  Scattering lengths: theory versus experiment. 
The shaded region represents the intersection of the 
domains allowed by the old data and by the Olsson sum rule, see
\protect{\cite{ACGL} for details.
The ellipse indicates the impact of the new, preliminary $K_{e_4}$
data \cite{Pislak}.  The cross shows the result of the 
present paper. The three diamonds illustrate the
convergence of the chiral perturbation series at threshold (see text).
The one at the left corresponds to Weinberg's leading order formulae. The
two black solid lines are the boundaries of the universal band
\protect{\cite{ACGL}}.\\

\vspace{0.7em}}
We emphasize that our result (\ref{final result}) relies on the standard
picture, according to which the quark condensate represents the leading
order parameter of the spontaneously broken symmetry.  The scenario
investigated in ref.~\cite{Stern Sazdjian Fuchs} concerns
the possibility that the Gell-Mann-Oakes Renner formula fails, the second
term in the expansion $ M_\pi^2= 2\, B\,m
\left\{1-\frac{1}{2}\,\xi\,\lbar_3 +O(\xi^2)\right\}$ being of the same
numerical order of magnitude or even larger than the first. Note that for
this to happen, the value of $|\lbar_3|$ must exceed the estimate
(\ref{l3barnum}) by more than an order of magnitude.

The constraints imposed on $a_0^0$, $a_0^2$ by the available experimental
information are shown in the figure. The ellipse represents the 68 \% 
confidence level contour obtained by combining the new, preliminary 
$K_{e_4}$ data \cite{Pislak} with earlier experimental results. 
Concerning the value of $a_0^0$, the ellipse
corresponds to the range $0.2<a_0^0<0.25$.  The representation
(\ref{l3rdependence}) shows that this range only puts very weak limits on
the value of $\lbar_3$. The precision data on the reaction $\pi N\rightarrow
\pi\pi N$ near threshold \cite{CHAOS} provide an independent measurement
of $a_0^0$. Unfortunately, however, the systematic errors of the 
Chew-Low extrapolation that underlies this
determination do not appear to be under good control \cite{Bolokhov} -- 
for a detailed discussion, we refer to \cite{pocanic}.   

The figure shows that the values of $a_0^2$ and $a_0^0$ are strongly
correlated. The correlation also manifests itself in the Olsson sum rule
\cite{Olsson}, which according to ref.~\cite{ACGL} leads to
$2a_0^0-5a_0^2=0.663 \pm0.021+1.13 \Delta a_0^0 -1.01\Delta a_0^2$, in
perfect agreement with our result in eq.~(\ref{final result}). Note,
however, that this combination is not sensitive to $\ell_3$ -- accurate
experimental information in the threshold region is needed to perform a
thorough test of the theoretical framework that underlies our
calculation. We are confident that the forthcoming results from Brookhaven
\cite{Pislak}, CERN \cite{DIRAC,cernkl4} and Frascati \cite{DAFNE} will
provide such a test.

\subsubsection*{Acknowledgement}
We thank I.~Mannelli and V.~Kekelidze for correspondence and for
pointing out to us ref.~\cite{cernkl4}. We enjoyed discussions with
S.~Pislak and thank him for providing us with the preliminary $K_{e4}$
data from the E865 experiment at Brookhaven. 
This work was supported by the Swiss National Science
Foundation, Contract No. 2000-55605.98, by TMR, BBW-Contract No. 97.0131  and
EC-Contract No. ERBFMRX-CT980169 (EURODA$\Phi$NE).

\end{document}